# Two-Dimensional Superconductivity at the $CaZrO_3/KTaO_3(001)$ Heterointerfaces

*Lu Chen,*[1,2,‡] *Siyi Zhou,*[3,‡] *Daming Tian,*[1,2] *Yinan Xiao,*[1,2] *Qixuan Gao,*[3] *Yongchao Wang,*[3] *Yuansha Chen,*[4,5] *Fengxia Hu,*[4,5] *Baogen Shen,*[4,5,6] *Jirong Sun,*[4,7,*] *Weisheng Zhao,*[1,2,*] *Jinsong Zhang,*[3,8,9,*] *and Hui Zhang*[1,2,*]

[1]School of Integrated Circuit Science and Engineering, Beihang University, Beijing 100191, China

[2]State Key Laboratory of Spintronics Hangzhou International Innovation Institute, Beihang University, Hangzhou 311115, China

[3]State Key Laboratory of Low Dimensional Quantum Physics, Department of Physics, Tsinghua University, Beijing 100084, China

[4]Beijing National Laboratory for Condensed Matter Physics and Institute of Physics, Chinese Academy of Sciences, Beijing 100190, China

[5]School of Physical Sciences, University of Chinese Academy of Sciences, Beijing 100049, China

[6]Ningbo Institute of Materials Technology & Engineering, Chinese Academy of Sciences,

Ningbo, Zhejiang, 315201, China

[7]School of Physics, Zhejiang University, Hangzhou 310027, China

[8]Frontier Science Center for Quantum Information, Beijing 100084, China

[9]Hefei National Laboratory, Hefei, Anhui, 230088, China

Keywords: two-dimensional superconductivity, oxide interface, two-dimensional electron gas, orientation-dependent, gating effect

Two-dimensional superconductivity at $KTaO_3$ (KTO) heterointerfaces has sparked intensive investigations since its discovery, yet whether the (001)-oriented KTO interface hosts superconductivity remains to be elucidated. Here, we provide unambiguous evidence of superconductivity in two-dimensional electron gases (2DEGs) at $CaZrO_3$/KTO(001) heterointerfaces, with a superconducting transition $T_C$ up to ~0.25 K. Notably, $T_C$ increases linearly with carrier density $n_S$ over the range of $4.5\times10^{13}$~$10.3\times10^{13}$ $cm^{-2}$. Furthermore, superconductivity exhibits a pronounced dependence on crystallographic orientation, with $T_C$ rising from 0.25 K for (001) to 1.04 K for (110) and 2.22 K for (111), underscoring the crucial role of interfacial symmetry in the $CaZrO_3$/KTO system. The two-dimensional nature of the superconducting state is corroborated by the Berezinskii-Kosterlitz-Thouless (BKT) transition and the large anisotropy of the upper critical field. For the $CaZrO_3$/KTO(001) sample with $n_S$=$7.7\times10^{13}$ $cm^{-2}$, the estimated Ginzburg-Landau coherence length $\xi_{GL}$=146.4 nm is larger than the superconducting layer thickness $d_{SC}$=10.1 nm by a factor of ~14.5, confirming significant two-dimensional confinement of the $CaZrO_3$/KTO(001) superconductor. In addition, we demonstrate that the two-dimensional superconductivity at the $CaZrO_3$/KTO(001) interface can be effectively tuned by applying a back gate voltage. Our findings reveal the existence of two-dimensional superconductivity at $CaZrO_3$/KTO(001), providing a new platform for exploring two-dimensional superconductivity at oxide interfaces.

## Introduction

Interfacial superconductivity has emerged as a central topic in quantum materials and condensed matter physics, providing a valuable platform for exploring pairing mechanisms.[1, 2] A prototypical example is the two-dimensional electron gas (2DEG) formed at the interface between the two band insulators $LaAlO_3$ (LAO) and $SrTiO_3$ (STO),[3] which hosts two-dimensional superconductivity with a transition temperature ($T_C$) of ~200 mK.[4-6] Remarkably, STO-based superconducting interfaces exhibit abundant intriguing quantum phenomena, including the coexistence of superconductivity and ferromagnetism,[7-9] as well as gate-tunable superconductivity.[10-12] More recently, the discovery of superconducting 2DEGs at $KTaO_3$ (KTO)-based interfaces has garnered considerable attention due to the substantially enhanced $T_C$ up to ~2 K at the KTO(111) interfaces,[13-16] nearly an order of magnitude higher than that observed at the LAO/STO interfaces. Although KTO shares many characteristics with STO, the 2DEGs residing in KTO exhibit significantly stronger spin-orbit coupling (SOC) due to the heavy Ta 5*d* orbitals at the heterointerface.[17-19] Strong SOC has been theoretically shown to play a crucial role in unconventional superconductivity by enabling mixed-parity pairing channels.[20, 21] In particular, strong SOC originating from Ta 5*d* orbitals has been reported to result in a mixed-parity superconducting state at the amorphous a-$YAlO_3$/KTO(111) interface, characterized by an admixture of *s*-wave and *p*-wave pairing components.[22]

Unlike the orientation-independent superconductivity observed in the STO-based 2DEGs,[4-6] the superconductivity state at KTO-based heterointerfaces exhibits a strong dependence on crystallographic orientation. Specifically, superconductivity emerges at both (111)- and (110)-oriented interfaces, with the maximal transition temperatures of 2.2 K[13] and 0.9 K,[23] respectively. In contrast, no superconductivity is detected at the KTO(001) interfaces down to 25 mK,[13] though metallic 2DEGs are also observed at this kind of interfaces. In fact, since the first report in 2011 of superconductivity on the KTO(001) surface gated by ionic liquid ($T_C$≈50 mK),[24] there has been no further progress in exploring superconductivity at KTO(001) surfaces or interfaces,[13, 25-29] despite extensive renewed interest sparked by the recent discovery of the superconductivity at KTO(111)-based heterointerfaces. Recently, several mechanisms have been proposed to account for the absence of a detectable superconducting transition at KTO(001) interfaces. One interpretation attributes this behavior to the large splitting between $d_{xy}$ and $d_{xz/yz}$ orbitals, which effectively decouples inter-orbital electrons from the soft transverse optical (TO1) phonons and suppresses superconductivity.[16] Another explanation, supported by soft X-ray angle-resolved photoemission spectroscopy (SX-ARPES) measurements of the interfacial bands, links the lack of superconductivity at KTO(001) interface to its weakest electron-phonon coupling among the three crystallographic orientations.[28] Unfortunately, these studies cannot answer the question of whether KTO(001) interfaces are superconductive or not. Obviously, the realization of superconducting KTO(001) interfaces is critical for establishing a complete chain of evidence for superconductivity in KTO-based interfaces, thereby enabling a comprehensive understanding of the underlying mechanism of interfacial superconductivity.

In this work, we resolve this central open question by demonstrating two-dimensional superconductivity in 2DEGs formed at the CZO/KTO(001) (CZO=$CaZrO_3$) heterointerfaces. The superconducting transition temperature $T_C$ exhibits a linear dependence on carrier density, reaching up to ~0.25 K for the (001) interface. In comparison, $T_C$ is 1.04 K and 2.22 K for the (110)- and (111)-oriented CZO/KTO interfaces, respectively. Thus, $T_C$ is lowest for (001), intermediate for (110), and highest for (111), indicating the strong influence of crystalline orientation on the superconducting properties of 2DEGs at the CZO/KTO interfaces. The Berezinskii-Kosterlitz-Thouless (BKT) transition analysis and the anisotropic upper critical field collectively confirm the two-dimensional nature of superconductivity at the CZO/KTO(001) interface. Moreover, electrical gate tuning enables an effective modulation of interfacial superconductivity. Our work not only demonstrates superconductivity at the (001)-oriented KTO heterointerface, offering a new platform for exploring emergent quantum phenomena in oxide heterostructures, but also provides valuable insights into the impact of crystalline orientation on interfacial superconductivity.

**Results and Discussion**

CZO thin films with thickness of 10 nm were grown on (001)-oriented KTO single-crystal substrates using pulsed laser deposition (PLD), following the procedure detailed in the Methods section. Atomic force microscopy (AFM) images indicate atomically flat surface morphologies for both the substrate and the film (Figure S1). The bulk lattice constant of CZO (pseudo-cubic $a_{CZO}$=4.012 Å) is very close to that of KTO ($a_{KTO}$=3.989 Å) with a lattice mismatch of as small as 0.58%, facilitating the layer-by-layer epitaxial growth of the CZO film. Due to the closely matched lattice constants, the diffraction peaks of the CZO film and KTO substrate nearly overlap, making them difficult to be resolved in $\theta$-2$\theta$ x-ray diffraction (XRD) pattern (Figure S2). This result is consistent with the previous report.[30] High-resolution scanning transmission electron microscopy (STEM) was performed to directly examine the crystal structure of the CZO(10 nm)/KTO(001) heterostructure, grown at a substrate temperature $T_S$=600 °C under an oxygen pressure of $P_{O2}$=2×10$^{-4}$ Pa. Figure 1a presents an atomically resolved high-angle annular dark-field (HAADF)-STEM image of the CZO(10 nm)/KTO(001) interface cross-section, recorded along the [010] zone axis of KTO. The large-scale STEM image and the corresponding Fast Fourier Transform (FFT) pattern are shown in Figure S3. Structural characterization indicates that the high-quality epitaxial growth of the crystalline CZO thin film on the KTO(001) substrate. Well-ordered arrays of A-site and B-site cations are observed in the HAADF image, with no visible dislocations or defects. Due to the bright contrast between Ta and Zr ions, an abrupt and smooth interface between the KTO substrate and the CZO layer can be easily identified. Line profile analysis of the atomic contrast along the vertical direction further indicates that the interface is atomically sharp, without interlayer mixing. Acquired from the same region as the HAADF-STEM image in Figure 1a, the energy dispersive X-ray spectroscopy (EDX) elemental mapping (Figure 1b and Figure S4) provides insight into the chemical composition across the interface, showing a clear interface without noticeable elemental interdiffusion. These results demonstrate a well-defined CZO/KTO(001) interface, both structurally and chemically.

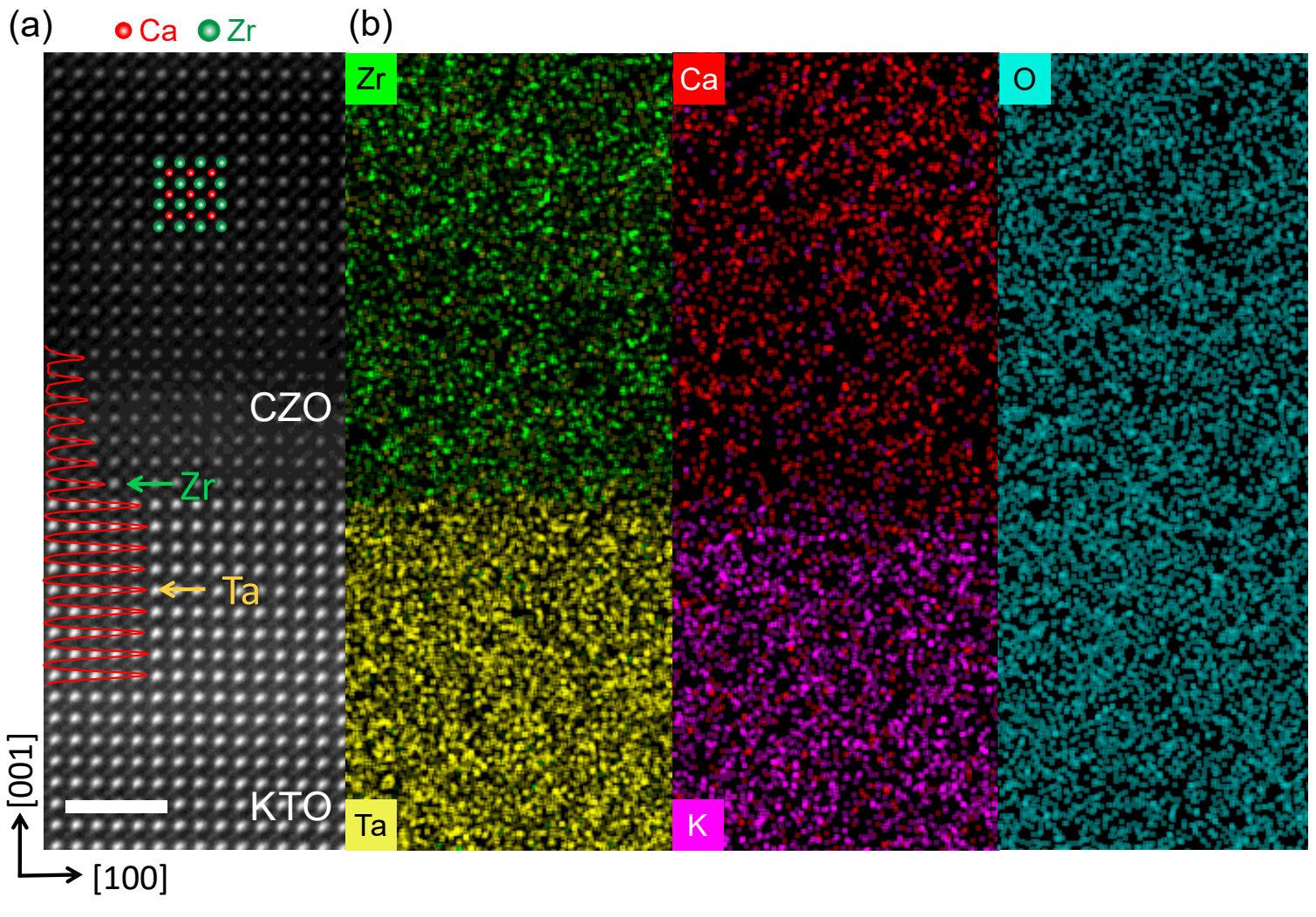

**Figure 1.** Epitaxial structure of the CZO/KTO(001) heterostructure. (a) High-angle annular dark-field (HAADF) image of the CZO(10 nm)/KTO(001) interface cross-section, recorded along the [010] zone axis. The scale bar represents 2 nm. The atomic configuration of CZO is schematically illustrated, with Ca and Zr lattice sites marked by red and green dots, respectively. The red curve is the vertical line profile of the HAADF image along the Ta/Zr atomic column. (b) Corresponding energy dispersive X-ray spectroscopy (EDX) elemental mapping of Zr, Ta, Ca, K and O across the interface.

The electrical transport properties of the CZO/KTO(001) heterointerfaces were measured using the Van der Pauw geometry. Metallic 2DEGs at the CZO/KTO(001) interfaces with varying sheet carrier densities ($n_S$) were obtained by adjusting $P_{O2}$ at a fixed substrate temperature ($T_S$=600 °C), as detailed in the Methods section and Table S1. Figure 2a displays the temperature dependence of the normalized sheet resistance ($R_S/R_{0.4K}$) for a series of CZO(10 nm)/KTO(001) samples with $n_S$ ranging from $4.5\times10^{13}$ to $10.3\times10^{13}$ cm$^{-2}$, where $R_{0.4K}$ denotes the normal-state resistance at 0.4 K. The unnormalized sheet resistance data are provided in Figure S5. These samples are labeled as Sample #, with # increasing sequentially with $n_S$. The carrier densities $n_S$ are determined from the magnetic-field-dependent Hall resistance $R_{XY}$ measured at $T$=2 K (Figure 2b), which increases as $P_{O2}$ decreases, suggesting that electron doping by oxygen vacancies plays an important role in forming the 2DEG. X-ray photoelectron spectroscopy (XPS) measurements further confirm the presence of $Ta^{4+}$ ions (Figure S6 and Table S2), indicating the formation of oxygen vacancies in the KTO substrate during CZO film growth. Notably, as shown in Figure 2a, the $R_S/R_{0.4K}$-$T$ curves for all samples drop sharply, undergoing a superconducting transition at low temperatures, indicating the universal nature and reproducibility of superconductivity at the CZO/KTO(001) interfaces. Meanwhile, the superconducting transition temperature $T_C$, defined as the temperature at which $R_S$ drops to 50% of the normal-state $R_{0.4K}$ value, is strongly dependent on carrier density, increasing progressively with $n_S$. As $n_S$ increases from $4.5\times10^{13}$ to $10.3\times10^{13}$ $cm^{-2}$, $T_C$ rises from 0.10 K to 0.25 K, exhibiting a more than twofold enhancement. In Figure 2c, we plot $T_C$ as a function of $n_S$, revealing a predominantly linear dependence with a slope of $\sim0.25\times10^{-14}$ $K\cdot cm^2$. This indicates that superconductivity is strongly affected by the carrier density, which is modulated by oxygen-vacancy-induced electron doping. A similar linear relationship between $T_C$ and $n_S$ has also been observed for the 2DEGs at the KTO(111) interfaces,[16, 31, 32] with a larger $T_C$- $n_S$ slope of $2\sim3\times10^{-14}$ $K\cdot cm^2$. This means that the rate at which $T_C$ increases with $n_S$ is significantly lower for the CZO/KTO(001) interfaces compared to the KTO(111) interfaces. By extrapolating the linear fit in Figure 2c, we obtain an intercept on the $n_S$ axis at $\sim0.76\times10^{13}$ $cm^{-2}$, indicating a critical carrier density below which superconductivity vanishes at the CZO/KTO(001) interface. Unlike earlier studies, where no superconducting transition was observed at KTO(001) interfaces down to ~25 mK,[13] our results reveal a robust superconducting transition at the CZO/KTO(001) interfaces. The maximum $T_C$ observed here is nearly five times higher than that of the electric double-layer gated KTO(001) surface[24] and is comparable to that of LAO/STO(001) interfaces.[4, 33]

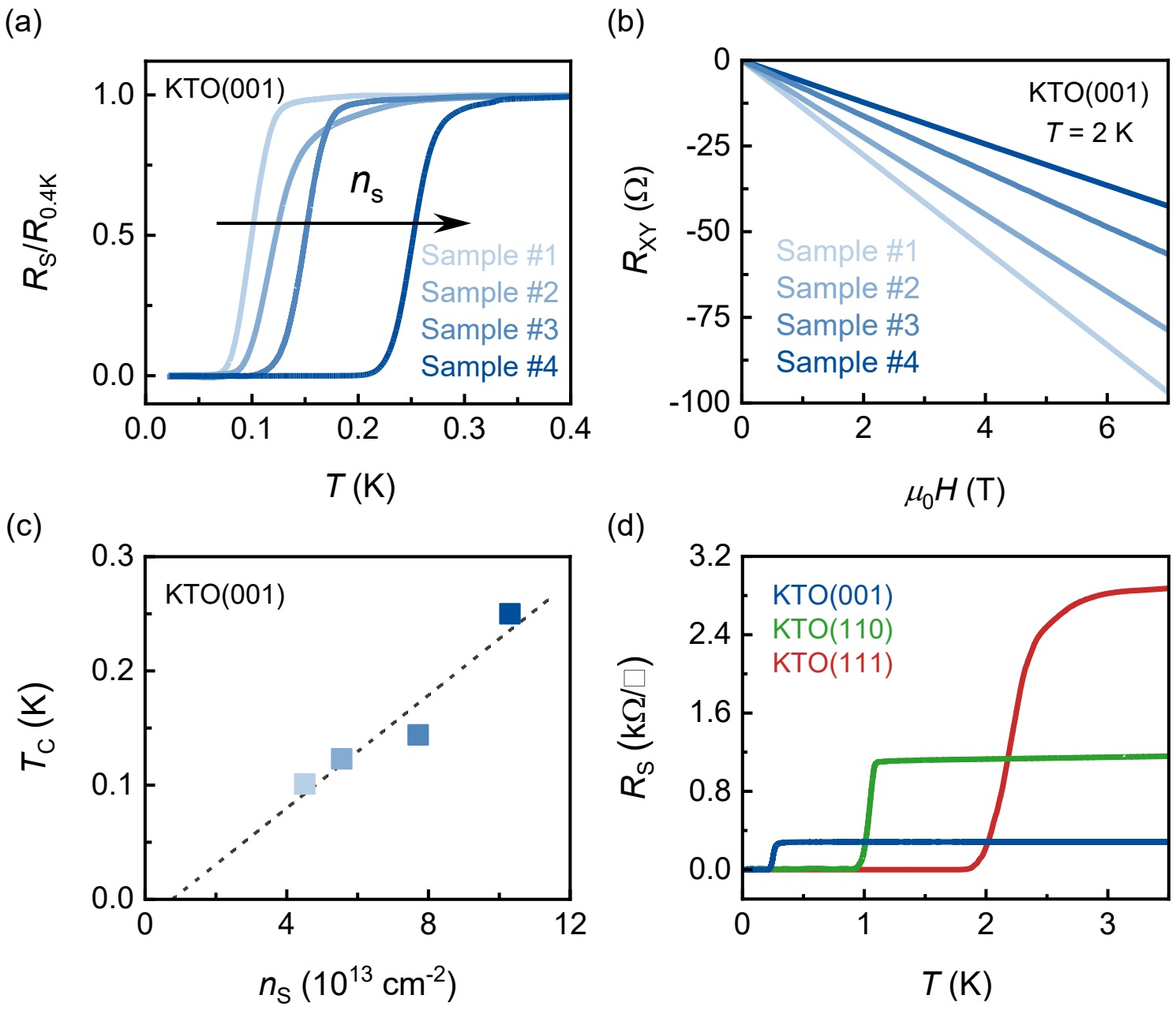

**Figure 2.** Superconducting transport properties of CZO/KTO heterostructures. (a) Normalized resistance $R_S/R_{0.4K}$ as a function of temperature for CZO/KTO(001) samples with different carrier densities $n_S$, where $R_{0.4K}$ is the normal-state resistance at 0.4 K. The direction of the arrow indicates the increase of $n_S$. (b) Hall resistance $R_{XY}$ as a function of magnetic field $\mu_0 H$ measured at 2 K. (c) Dependence of the superconducting transition temperature $T_C$ on $n_S$ for (001)-oriented CZO/KTO samples. (d) Temperature dependence of the sheet resistance $R_S$ for CZO/KTO(001), (110) and (111) heterostructures prepared under identical growth conditions.

To further investigate the dependence of interfacial superconductivity on crystallographic orientation, reference CZO(10 nm)/KTO samples with (110) and (111) orientations were prepared under the same growth conditions as the (001)-oriented sample #4. All three samples exhibit globally metallic behavior over a wide temperature range (Figure S7), confirming the formation of 2DEGs at their interfaces. We compare the temperature dependence of $R_S$ in the low-temperature range for the 2DEGs formed at the (001)-, (110)- and (111)-oriented CZO/KTO interfaces in Figure 2d. All three samples undergo a transition to the superconducting state as the temperature decreases, with $T_C$ exhibiting a strong dependence on interfacial orientation. As the orientation varies from (001) to (110) and (111), $T_C$ increases from 0.25 K to 1.04 K and 2.22 K, following the trend: $T_C$ (001)< $T_C$ (110)< $T_C$ (111). The $T_C$ of the (001)-oriented interface is nearly an order of magnitude lower than that of the (111)-oriented interface, indicating the significant impact of crystal orientation on superconductivity at the CZO/KTO interfaces. Note that the $T_C$ values for the (110) and (111) orientations in our work are consistent with previous reports.[13, 23]

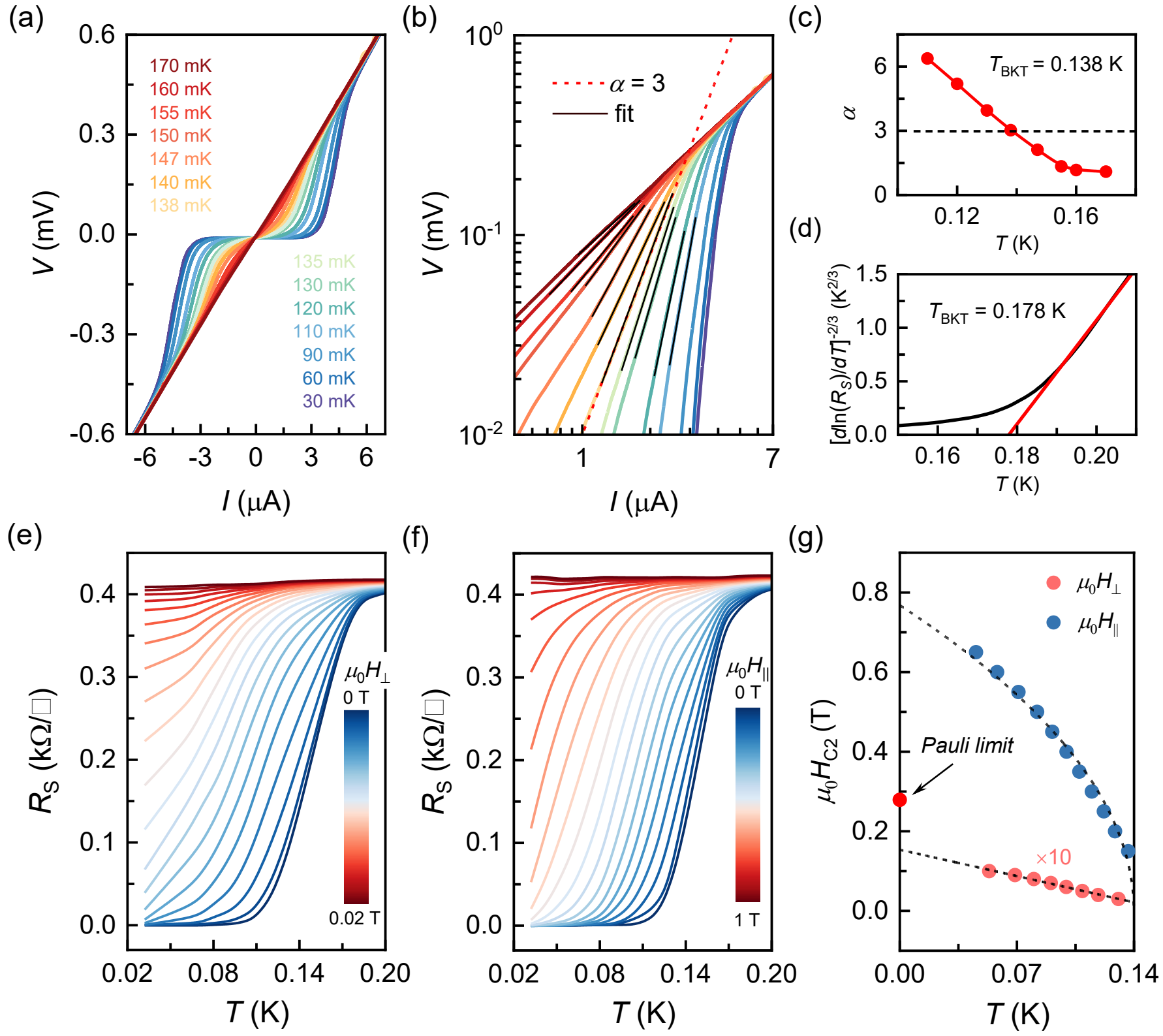

**Figure 3.** Berezinskii-Kosterlitz-Thouless (BKT) transition of CZO/KTO(001). (a) Temperature-dependent *I-V* measurements. (b) *I-V* curves plotted on a logarithmic scale, using the same symbol label as in a). Black solid lines are linear fits to the data. The red dashed line corresponds to $V \propto I^3$, which is used to infer the BKT transition temperature. (c) Temperature dependence of the power-law exponent $\alpha$, deduced from the linear fits in (b). (d) Sheet resistance dependence of temperature plotted on a $[d\ln(R_S)/dT]^{-2/3}$ scale. The red line is a linear extrapolation from the high-$T$ linear

section, which crosses the $T$-axis at $T_{BKT}$=0.178 K. Sheet resistivity $R_S$ as a function of temperature, measured under (e) out-of-plane magnetic fields from 0 to 0.02 T, stepped by 1 mT, and (f) in-plane magnetic fields from 0 to 1 T, stepped by 0.05 T. (g) Temperature dependence of the upper critical field $\mu_0 H_{C2}$, extracted from the 50% normal-state resistance at 0.2 K. Dashed lines are fitting curves based on Ginzburg-Landau theory. The estimated Pauli paramagnetic limit is marked with a red dot, which is significantly surpassed by $\mu_0 H_{C2||}(0)$.

To gain further insight into the superconductivity at the CZO/KTO(001) interfaces, we measured the current-voltage ($I$-$V$) characteristics of Sample #3 over a temperature range from 30 mK to 170 mK, as shown in Figure 3a. Below $T_C$, a robust critical-current behavior is observed, further confirming the existence of superconductivity at the interface. The critical current $I_C$ increases systematically with decreasing temperature, reaching a maximum of ~3.1 μA at $T$=30 mK. For a two-dimensional superconductor, the Berezinskii-Kosterlitz-Thouless (BKT) transition can be characterized by a $V \propto I^{\alpha}$ power-law relation with $\alpha$=3 at the transition temperature $T_{BKT}$, corresponding to the transition from unpaired vortex and antivortex to bound vortex-antivortex pairs.[4, 34] In Figure 3b, the $I$-$V$ curves are plotted on a log-log scale, where the red dashed line represents an $I^3$ dependence. The temperature dependence of the power-law exponent $\alpha$, extracted from the linear fits (black solid lines in Figure 3b), is presented in Figure 3c, from which $T_{BKT}$=0.138 K is determined. In addition, $T_{BKT}$ can also be estimated using the formula $R_S(T)=R_0\exp[-b(T/T_{BKT}-1)^{-1/2}]$, where $R_0$ and $b$ are material parameters.[35] As illustrated in Figure 3d, fitting this relation to the measured $R_S(T)$ yields $T_{BKT}$=0.178 K, consistent with the value derived from the analysis of the $I$-$V$ characteristics. These findings confirm the two-dimensional nature of the superconductivity at the CZO/KTO(001) interfaces.

To investigate the effect of the magnetic field on superconducting properties, we systematically measured the temperature-dependent sheet resistance $R_S$ under out-of-plane ($\mu_0 H_\perp$) and in-plane ($\mu_0 H_\parallel$) magnetic fields, as shown in Figures 3e and 3f, respectively. Here, $\mu_0$ is the vacuum permeability. The temperature-dependent upper critical fields $\mu_0 H_{C2\perp}$ and $\mu_0 H_{C2\parallel}$, extracted at the midpoints of the normal-state resistance at 0.2 K from the $R_S$-$T$ curves in Figures 3e and 3f, are summarized in Figure 3g. We further performed an analysis of the $\mu_0 H_{C2}$-$T$ relation based on the Ginzburg-Landau theory[36]:

$$\mu_0 H_{C2\perp}(T)=[\phi_0/2\pi\xi_{GL}^2](1-T/T_C) \tag{1}$$

$$\mu_0 H_{C2\parallel}(T)=[\phi_0\sqrt{12}/2\pi\xi_{GL}d_{SC}](1-T/T_C)^{\frac{1}{2}} \tag{2}$$

where $\phi_0$ is the flux quantum, $\xi_{GL}$ is the Ginzburg-Landau coherence length at $T$=0 K, and $d_{SC}$ is the superconducting layer thickness. The fits to Eqs. (1) and (2) yield the extrapolated zero-temperature critical fields $\mu_0 H_{C2\perp}(0)$=0.015 T and $\mu_0 H_{C2\parallel}(0)$=0.767 T, corresponding to a large anisotropic ratio of ~51. The deduced coherence length $\xi_{GL}$=146.4 nm exceeds the superconducting layer thickness $d_{SC}$=10.1 nm by a factor of ~14.5, demonstrating significant two-dimensional confinement of the CZO/KTO(001) superconductor. We summarize previous studies on the superconducting parameters of 2DEGs at typical oxide heterointerfaces, as outlined in Table 1. A comparison of KTO-based systems reveals that although the $T_C$ of the (001)-oriented KTO interface is the lowest—nearly an order of magnitude smaller than that of the (111) system—it nevertheless exhibits an exceptionally long Ginzburg-Landau coherence length ($\xi_{GL}$≈146 nm), far exceeding the typical values reported for the KTO(110) and (111) interfaces ($\xi_{GL}$≈13~27 nm). In addition, the ratio $\xi_{GL}/d_{SC}$ is substantially higher than that of the (110) and (111) systems, indicating the markedly more pronounced two-dimensional character of the superconducting state at the CZO/KTO(001) interface. This suggests that the superconducting order parameter in the CZO/KTO(001) system is more tightly confined to the two-dimensional plane, which may provide a favorable platform for exploring superconducting pairing, tuning superconductivity, and uncovering emergent quantum phases in oxide heterostructures.

The mean free path of the conducting electrons can be estimated as $l_{\mathrm{mfp}}=(h/\mathrm{e}^2)(1/k_{\mathrm{F}}R_{\mathrm{S}})$ in a single band model, where $h$ is Planck constant and $k_{\mathrm{F}}=(2\pi n_{\mathrm{S}})^{1/2}$ is the Fermi wave number.[37] Using the $n_{\mathrm{S}}$ and $R_{\mathrm{S}}$ values measured at $T$=2 K, we obtain $l_{\mathrm{mfp}}$=25.1 nm, which is approximately 17.1% of $\xi_{\mathrm{GL}}$. Therefore, the superconductivity at the CZO/KTO(001) interface occurs in an intermediate range between clean ($l_{\mathrm{mfp}} \gg \xi_{\mathrm{GL}}$) and dirty ($l_{\mathrm{mfp}} \ll \xi_{\mathrm{GL}}$) limits, which is similar to a-LAO/KTO(110)[23] and EuO/KTO(111).[13]

**Table 1.** Comparison of superconducting parameters in 2DEGs at oxide heterointerfaces.

| **Heterostructures** | **Orientations** | $\boldsymbol{T_{\mathrm{C}}}$ **(K)** | $\boldsymbol{\xi_{\mathrm{GL}}}$ **(nm)** | $\boldsymbol{d_{\mathrm{SC}}}$ **(nm)** | $\boldsymbol{\xi_{\mathrm{GL}}/d_{\mathrm{SC}}}$ |
|---|---|---|---|---|---|
| CZO/KTO (this work) | (001) | 0.15 | 146.4 | 10.1 | 14.5 |
| a-LAO/KTO[14, 23] | (110) | 0.93 | 27.3 | 8.0 | 3.4 |
| | (111) | 2.00 | 18.8 | 4.0 | 4.7 |
| EuO/KTO[13, 18] | (110) | 1.35 | 27.0 | 6.4 | 4.2 |
| | (111) | 1.74 | 13.0 | 5.1 | 2.5 |
| a-$\mathrm{AlO_x}$/KTO[31] | (111) | 1.55 | 23.4 | 5.0 | 4.7 |
| a-$\mathrm{YAlO_3}$/KTO[22] | (111) | 1.86 | 18.4 | 4.5 | 4.1 |
| LAO/STO[33, 38, 39] | (001) | 0.20 | 70.0 | 10.0 | 7.0 |
| | (110) | 0.15 | 44.0 | 24.0 | 1.8 |
| | (111) | 0.30 | 26.0 | 7.0 | 3.7 |

For weakly coupled BCS superconductors, the Pauli paramagnetic limit set the theoretical upper bound for the in-plane critical field[40, 41], given by $\mu_0 H_{\mathrm{C2}}^{\mathrm{P}} \approx 1.76 k_{\mathrm{B}} T_{\mathrm{C}}/\sqrt{2}\mu_{\mathrm{B}}$, where $k_{\mathrm{B}}$ and $\mu_{\mathrm{B}}$ are the Boltzmann constant and Bohr magneton, respectively. Using $T_{\mathrm{C}}$=0.15 K, we obtain $\mu_0 H_{\mathrm{C2}}^{\mathrm{P}}$=0.28 T (denoted by a red solid circle in Figure 3g), which is only 36% of $\mu_0 H_{C2\parallel}(0)$=0.77 T. Notably, the Pauli paramagnetic limit is significantly exceeded, as previously observed in EuO/KTO(110),[42] which could be attributed to the strong SOC at the 5*d* KTO interfaces. As mentioned above, the strong spin-orbit coupling is expected to give rise to unconventional superconductivity, suggesting that the KTO(001) interface likely hosts unconventional pairing.

To explore the tunability of superconductivity via electric field, we performed electrostatic gating measurements. A schematic of the field-effect device is provided in the inset of Figure 4a. As illustrated in Figure 4a, superconductivity can be effectively tuned by applying a back-gate voltage ($V_G$) across the KTO substrate, as evidenced by the systematic evolution of the temperature-dependent $R_S$ under different $V_G$ values. The normal-state sheet resistance $R_S$(0.4 K) increases monotonically from 300 to 483 Ω/□ when $V_G$ is varied from +150 V to −150 V, while the superconducting transition shifts first to higher temperatures and then to lower temperatures. To characterize the gate-tunable transport properties, we carried out normal-state Hall effect measurements at 2 K under various $V_G$ values. The extracted carrier density $n_S$ and Hall mobility $\mu$ are summarized in Figure 4b. As $V_G$ changes from +150 V to –150 V, $n_S$ decreases from ~$8.1\times10^{13}$ to ~$7.1\times10^{13}$ $cm^{-2}$, while $\mu$ decreases from ~166 to ~126 $cm^2$/Vs. Applying a negative $V_G$ depletes charge carriers and simultaneously enhances their spatial confinement near the CZO/KTO(001) interface, where the higher degree of disorder leads to a reduction in Hall mobility. The increased disorder is further evidenced by the decrease in the mean free path ($l_{mfp}$) with decreasing $V_G$ (Figure S8). Notably, as shown in Figure 4c, the mid-point $T_C$ extracted from Figure 4a exhibits a dome-shaped dependence on $V_G$, peaking at $V_G$=+30 V. As illustrated in Figure 4d, the critical current $I_C$ can also be tuned by changing the gate voltage $V_G$ at $T$=30 mK. Notably, $I_C$ exhibits a similar non-monotonical dependence on $V_G$, first increasing and then decreasing as $V_G$ is swept from +150 V to –150 V. Similar dome-shaped superconducting phase diagrams under electrostatic gating have been observed in other oxide 2DEG systems.[10, 14, 24]

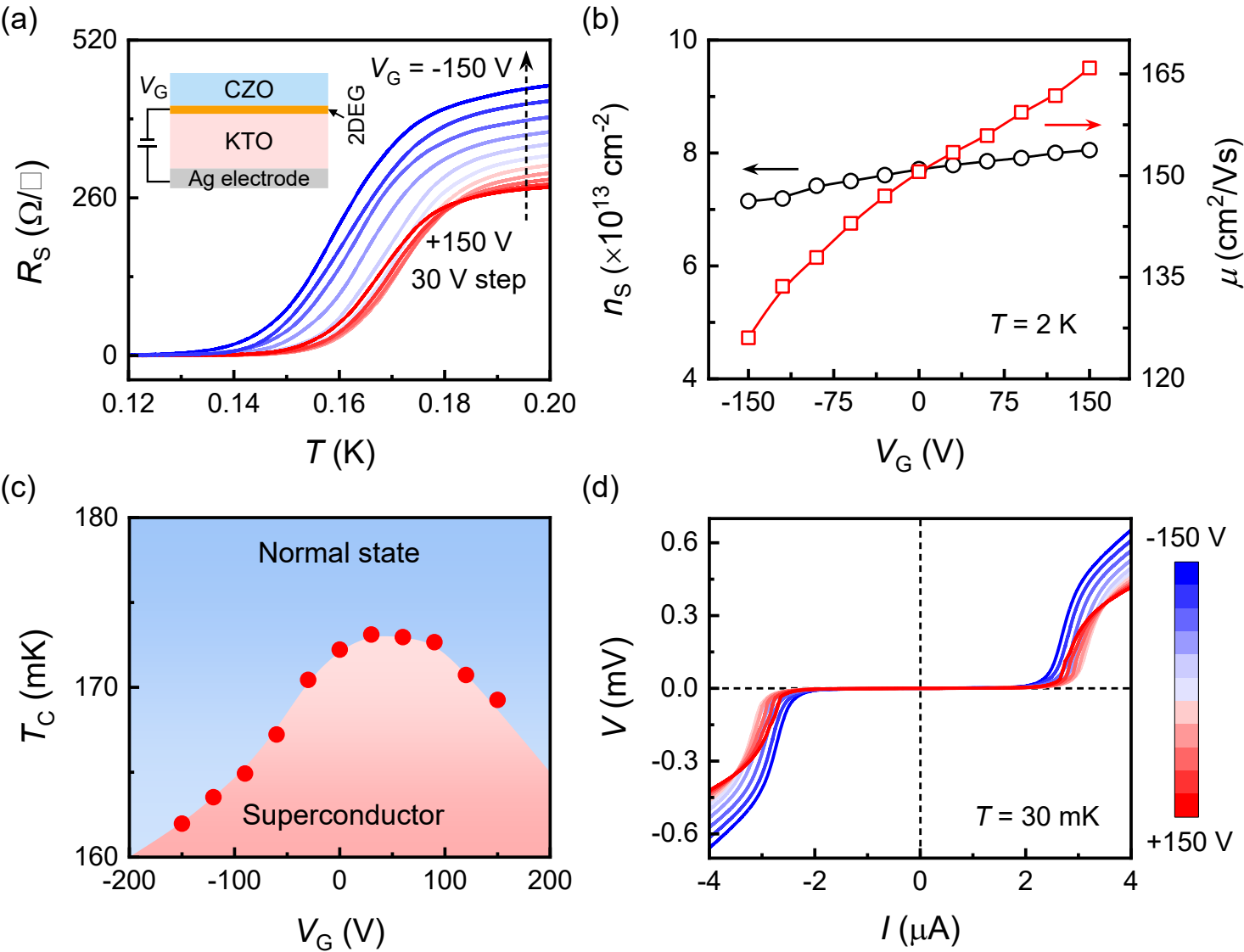

**Figure 4.** Electric field control of transport properties. (a) Temperature dependence of the sheet resistance $R_S$ under back-gate voltages $V_G$ ranging from +150 V to –150 V in steps of 30 V. Inset: Schematic of the back gate measurement geometry. (b) Carrier density $n_S$ and Hall mobility $\mu$ as functions of $V_G$ at 2 K. (c) Midpoint $T_C$ plotted against $V_G$, revealing the superconducting dome in the phase diagram. (d) *I-V* measurement at different $V_G$ values at $T$=30 mK.

To elucidate why superconductivity emerges at CZO/KTO(001) interfaces while being absent in previous studies of KTO(001) heterointerfaces, we first examine whether differences in carrier density could account for this discrepancy. A systematic review of earlier reports on KTO(001)-based heterointerfaces reveals that the carrier densities previously achieved are comparable to those in our superconducting CZO/KTO(001) samples (Figure 5a and Table S3). This indicates that carrier density alone is unlikely to account for the absence of superconductivity in earlier studies. It instead points to the importance of interfacial properties, with the overlayer possibly playing the critical role in the emergence of superconductivity at the CZO/KTO(001) interface.

We further fabricated CZO(10 nm)/KTO(001) samples at varying substrate temperatures $T_S$ with a fixed oxygen pressure of $P_{O2}$=2×10$^{-4}$ Pa. Notably, as shown in Figure 5b, superconductivity is absent in samples grown at lower temperatures, despite the formation of 2DEGs in these cases. Figure 5c shows the dependence of $T_C$ on the substrate temperature. The epitaxial quality of the CZO film is strongly influenced by the growth conditions, particularly the substrate temperature. As reported previously,[30] the reflected high-energy electron diffraction (RHEED) patterns of the CZO film vanish at deposition temperatures of 400 °C or below, indicating that the CZO film becomes disordered or amorphous. As the growth temperature decreases, the CZO film gradually evolves from a crystalline to an amorphous state, which may modify the interfacial polar and electronic reconstruction, thereby affecting the superconducting properties. Our results highlight that the crystallinity of the CZO overlayer is a key factor for superconductivity at the CZO/KTO(001) interface, suggesting that this overlayer may provide unique advantages, the precise origins of which require further investigation.

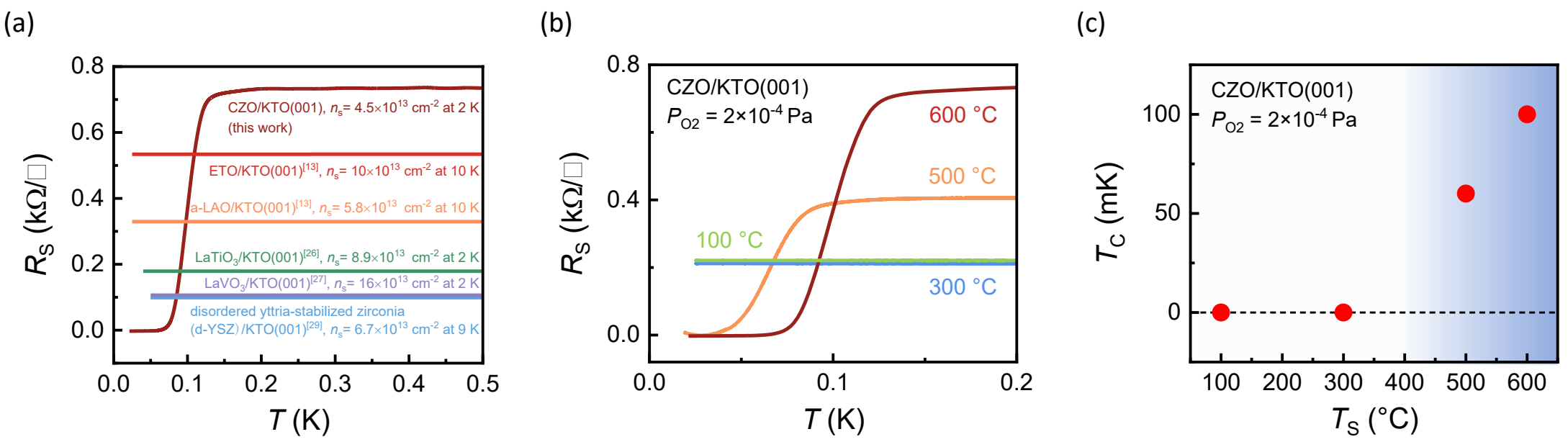

**Figure 5.** (a) Temperature dependence of the sheet resistance $R_S$ for our CZO/KTO(001) sample with $n_S$=4.5×10$^{-13}$ cm$^{-2}$, compared with preciously reported representative data for EuO/KTO(001),[13] a-LAO/KTO(001),[13] $LaTiO_3$/KTO(001),[26] $LaVO_3$/KTO(001),[27] and disordered yttria-stabilized zirconia (d-YSZ)/KTO(001).[29] (b) $R_S$ as a function of temperature for CZO/KTO(001) samples grown at different substrate temperatures with a fixed oxygen pressure

of $P_{O2}$=2×10$^{-4}$ Pa. (c) Dependence of $T_C$ on the substrate temperature, summarized from the data in (b).

## Conclusion

In summary, we report the discovery of two-dimensional superconductivity in 2DEGs formed at the interfaces between CZO films and (001)-oriented KTO single-crystal substrates. The superconducting transition temperature $T_C$ at the (001) interface reaches up to ~0.25 K and exhibits a linear dependence on carrier density. Note that this value is significantly lower than the $T_C$ values of 1.04 K and 2.22 K observed for the CZO/KTO(110) and (111) interfaces, respectively, underscoring the strong influence of crystalline orientation on interfacial superconductivity. This interfacial superconductivity is confirmed to be two-dimensional in nature, as evidenced by the BKT transition behavior and the large anisotropy of the upper critical field. The superconducting layer thickness and the coherence length are estimated to be ~10.1 and ~146.4 nm, respectively, indicating an exceptionally strong two-dimensional confinement. Additionally, the interfacial superconductivity can be effectively modulated through electrostatic gating. These findings expand the research framework for two-dimensional superconductivity at oxide interfaces, revealing a new platform for exploring superconducting quantum phenomena.

## Methods

Sample fabrication: CZO films with thickness of 10 nm were grown on 5×5 mm$^2$ KTO single-crystalline substrates with (001), (110), and (111) orientations by pulsed laser deposition (PLD) with a ceramic $CaZrO_3$ target. The KrF laser (wavelength is 248 nm) was set to a repetition rate of 2 Hz and a fluence of ~2 J/cm$^2$. The first set of samples was prepared by setting substrate temperature $T_S$ at 600 °C, but varying oxygen partial pressure $P_{O2}$ from 2×10$^{-4}$ to 3×10$^{-5}$ Pa. The second set of samples was fabricated under different $T_S$ from 600 °C down to 100 °C, with a fixed oxygen pressure of $P_{O2}$=2×10$^{-4}$ Pa. After deposition, the samples were cooled to room temperature in the same oxygen pressure environment. Surface morphology was analyzed using an atomic force microscope (AFM, MultiMode 8, Bruker). Film thickness was determined by the number of laser pulses, calibrated via small-angle X-ray reflectivity. The crystal structure of the films was characterized by $\theta$-2$\theta$ x-ray diffraction (XRD) in a high-resolution mode using Cu Kα emission (wavelength $\lambda$=1.5418 Å, Bruker D8 Discover). Cross-sectional specimens for electron microscopy investigations were prepared using a Focused Ion Beam (FIB) (Nova 200 NanoLab). Lattice images were obtained using a high-resolution scanning transmission electron microscope (STEM) with double $C_S$ correctors (JEOL-ARM200F). X-ray photoelectron spectroscopy (XPS) measurements were conducted using an Al Kα source in a ThermoFisher Scientific ESCALAB 250X.

Electrical transport measurements: Electrical transport measurements were measured using the Van der Pauw geometry. Ultrasonic wire bonding with 20 μm diameter Al wires was employed for electrical connections. Measurements of $R_S$-$T$ and Hall effect were performed in a Quantum Design Physical Property Measurement System (PPMS) over a temperature range of 300 K to 2 K, with a constant current of 10 μA applied. The electrical transport measurements at lower temperatures were performed in a dilution refrigerator equipped with a vector-type magnet, applying a 10 nA AC current at a frequency of 17.7777 Hz (Keithley 6221 current source). The corresponding AC voltage signals were recorded using NF LI5650 lock-in amplifiers. For $I$-$V$ curve measurements, a DC current was applied using the Keithley 6221 current source, and the corresponding voltage was measured with a Keithley 2182 Nanovoltmeter.

ASSOCIATED CONTENT

A preprint version of this manuscript is available online:

L. Chen; S. Zhou; D. Tian; etc. Two-Dimensional Superconductivity at the $CaZrO_3/KTaO_3$ (001) Heterointerfaces. 2025, arXiv:2507.01392. arXiv. https://doi.org/10.48550/arXiv.2507.01392 (accessed July 2, 2025).

**Supporting Information**.

Details of the procedures for 2DEGs fabricated at the CZO/KTO(001) interfaces; Summary of transport properties of KTO (001)-based 2DEGs from recent works; AFM images of thin films and KTO substrates; XRD images of thin films and KTO substrates; Large-scale HAADF-STEM image and Fast Fourier Transform (FFT) pattern; EDX elemental mapping; XPS of Ta 4*f* for the CZO/KTO(001) heterostructure; Transport properties of the (001)-, (110)-, and (111)-oriented CZO/KTO samples; Mean free path $l_{mfp}$ under back-gate voltages $V_G$.

**Corresponding Author**

Hui Zhang - School of Integrated Circuit Science and Engineering, Beihang University, Beijing 100191, China; State Key Laboratory of Spintronics Hangzhou International Innovation Institute, Beihang University, Hangzhou 311115, China; Email: huizh@buaa.edu.ac.cn

Jirong Sun - Beijing National Laboratory for Condensed Matter Physics and Institute of Physics, Chinese Academy of Sciences, Beijing 100190, China; School of Physics, Zhejiang University, Hangzhou 310027, China Email: jrsun@iphy.ac.cn

Weisheng Zhao - School of Integrated Circuit Science and Engineering, Beihang University, Beijing 100191, China; State Key Laboratory of Spintronics Hangzhou International Innovation Institute, Beihang University, Hangzhou 311115, China; Email: wszhao@buaa.edu.cn

Jinsong Zhang - State Key Laboratory of Low Dimensional Quantum Physics, Department of Physics, Tsinghua University, Beijing 100084, China; Frontier Science Center for Quantum Information, Beijing 100084, China; Hefei National Laboratory, Hefei, Anhui, 230088, China; Email: jinsongzhang@mail.tsinghua.edu.cn

**Author Contributions**

H. Z. conceived the research project. H. Z. J. R. S. W. S. Z. and J. S. Z. designed the experiments; L. C. Y. N. X. and D. M. T. prepared the samples; L. C. S. Y. Z. Y. N. X. Q. X. G. and Y. C. W. performed the transport measurements; H. Z. L. C. Y. S. C. F. X. H. B. G. S. and J. R. S. analyzed the data; H. Z. L. C. and J. R. S. wrote the manuscript with input from all authors. ‡These authors contributed equally.

**Notes**

The authors declare no competing financial interest.

**Acknowledgments**

This work has been supported by the Science Center of the National Science Foundation of China (Grant No. 52088101, and No. 52388201), the National Key Research and Development Program of China (Grant No. 2022YFA1403302, No. 2025YFA1411003, No. 2024YFA1410200, No. 2023YFA1406003, No. 2021YFA1400300, and No. 2024YFA1409100), the National Natural Science Foundation of China (Grant No. 12474103, No. T2394474, No. 12274443, No. 12350404,

No. 12274252, No. 92263202, No. U23A20550, No. 22361132534, and No. U25A6010), the Strategic Priority Research Program B of the Chinese Academy of Sciences (Grant No. XDB1270201), the Innovation Program for Quantum Science and Technology (Grant No. 2021ZD0302502). W.S.Z is thankful for the support of the Beijing Outstanding Young Scientist Program.